# Low Temperature Formation of Crystalline $VO_2$ Domains in Porous Nanocolumnar Thin Films for Thermochromic Applications

H. Acosta-Rivera[1], V. Rico[1*], F.J. Ferrer[2,3], T.C. Rojas[1], R. Alvarez[1,4], N. Martin[5], A.R. Gonzalez-Elipe[1], A. Palmero[1*]

[1] Instituto de Ciencia de Materiales de Sevilla (CSIC-US), Américo Vespucio 49, E-41092 Seville, Spain

[2] Centro Nacional de Aceleradores (CSIC-US), Thomas A. Edison 7, E-41092, Seville, Spain

[3] Departamento de Física Atómica, Molecular y Nuclear, Universidad de Sevilla, Aptdo 1065, E-41012 Seville, Spain

[4] Departamento de Física Aplicada I. Escuela Politécnica Superior, Universidad de Sevilla, Virgen de África 7, E-41011 Seville, Spain

[5] SUPMICROTECH, CNRS, Institut FEMTO-ST, F-25000 Besançon Cedex, France

**Abstract**.- The formation of $VO_2$ crystalline domains in amorphous substoichiometric nanocolumnar $VO_x$ thin films subjected to an oxidation process at temperatures below 300ºC has been studied. It is obtained that values of [O]/[V] above 1.9 lead to the sole formation of $V_2O_5$ after oxidation, while values below 1.9 favor the formation of $VO_2$, $V_3O_7$ and $V_2O_5$ crystalline domains for temperatures as low as 260ºC. Moreover, it is found that the adsorption of oxygen and its incorporation into the film network produce a relevant volume expansion in a so-called swelling mechanism that makes pores shrink. Under some specific conditions, the low temperature oxidation does not only trigger the formation of $VO_2$ domains but also a drastic reduction of oxygen-deficient amorphous $VO_x$ in the films, which clearly improves the overall transparency and thermochromic modulation capabilities. The changes in the optical and electrical properties of these films during the metal-insulator transition have been studied, finding the best performance when the stoichiometry of the film before oxidation is [O]/[V]=1.5 and the oxidation temperature 280ºC. These conditions yield a relatively transparent coating that presents an optical modulation in the near-infrared range of nearly 50% and a drop of electrical resistivity of more than two orders of magnitude. A tentative model based on the volume increase experienced by film upon oxidation is proposed to link the structural/chemical features of the films and the formation of $VO_2$ domains at such relatively low temperatures.

* Email: victor@icmse.csic.es; alberto.palmero@csic.es

## Introduction

$VO_2$ is a well-known material that undergoes a Metal-Insulator Transition (MIT) at near room temperature, nominally at $T_f$ = 68ºC, which induces a shift in its crystalline structure from monoclinic to rutile and causes a drastic change in its electrical properties [1]. This feature has motivated the use of $VO_2$ for numerous applications in different fields, e.g., as a channel layer in field-effect transistors (FETs), in memory devices, or in strain and gas sensor devices, among others (see for instance [2] and references therein). Additionally, the MIT causes a notable change in the optical transmittance of the material, particularly in the near-IR part of the spectrum, which also makes $VO_2$ suitable for solar irradiation control applications in interior spaces and buildings, commonly known as "smart windows" applications [3]. Here, the core functional principle relies on the ability to use $VO_2$-coated glazing to filter out near-IR light when the environmental temperature exceeds $T_f$, while maintaining a relatively high transparency in the visible range [1, 4]. Since approximately 52% of sunlight at sea level falls within the IR region, such glazing would enable a passive and indirect control of the temperature in interiors, reducing the use of alternative energy-intensive environmental control approaches, such as air conditioning or heating systems. Yet, the use of $VO_2$ for smart windows applications is currently hindered by several scientific and technical constraints. These include the need to lower $T_f$ to values closer to human comfort levels, ideally between 20 and 40ºC, as well as the optimization of the optical modulation in the near-IR region of the spectrum and of the transparency of the glazing in the visible range [3, 5]. While these issues are being actively studied in the literature, an additional critical constraint relies on the very fabrication method, as current procedures demand processing temperatures well above 500ºC to trigger the crystallization of $VO_2$ in its monoclinic phase, thus discarding the use of numerous and common temperature sensitive materials to support the coating [6]. In this paper we focus on this last issue, and study a basic and simple strategy to trigger the formation and crystallization of $VO_2$ at temperatures below 300ºC in porous thin films.

Several strategies are currently being explored in the literature to achieve the formation of monoclinic $VO_2$ in thin films at low temperatures. These typically rely on a multistep technique, where a V-O-containing thin film is first synthesized and, subsequently, exposed to an oxidizing or reductive atmosphere at relatively low temperatures under the influence of an additional excitation source, such as lasers or UV lights [6-9]. For instance, in ref. [6], $VO_2$ thin films fabricated via the sol-gel technique were treated with an excimer laser, achieving laser-induced crystallization within 60 seconds of exposure. Moreover, in ref. [7], the annealing temperature was reduced to approximately 250ºC by combining deep ultraviolet irradiation with a carbon-free oxidizer to generate reactive radicals. A distinct approach was followed in ref. [10], in which a very thin bilayer structure made of V and $V_2O_5$ was annealed in an oxygen atmosphere at temperatures below 300ºC. This procedure resulted in the formation of monoclinic $VO_2$ crystallites with thermochromic activity, thus demonstrating that this treatment at such relatively low temperatures may trigger the crystallization of $VO_2$. Following this line of reasoning, our group has recently demonstrated that the low temperature oxidation of a highly porous and amorphous nanocolumnar $VO_x$ thin film could also induce the formation of $VO_2$ crystallites at temperatures as low as 270ºC [11]. This preliminary result suggests that, upon optimization, this approach may serve as a basis for the low temperature fabrication of monoclinic $VO_2$ thin films. This is the main scope of this paper, in which the effect of low temperature oxidation of highly porous and amorphous $VO_x$ thin films is systematically analyzed in connection with the formation of monoclinic $VO_2$ crystal domains and the features of the metal-insulator transition.

Currently, there are numerous works that analyze the features and properties of porous nanocolumnar thin films grown by Physical Vapor Deposition techniques, such as magnetron sputtering [12] or electron beam-assisted evaporation [13]. Specifically, the so-called magnetron sputtering technique at oblique angles (MS-OAD) has been developed in the last decade as a variation of the classical magnetron sputtering deposition technique, by which a low pressure plasma is ignited in a vacuum reactor to sputter atoms from a solid target, that are subsequently deposited on a substrate, making the film grow [14]. In this way, the MS-OAD technique proposes the use of a particular geometrical arrangement in the vacuum reactor by which the substrate is tilted with respect to the target in order to induce the oblique incidence of the deposition species [15]. This results in the formation of tilted and randomly scattered nanocolumnar structures, approximately a hundred nanometers wide, separated by large accessible mesopores that extend from the top surface of the film to the very substrate. Moreover, these nanocolumnar structures present a certain degree of internal open porosity in the form of micropores which, being connected to the large intercolumnar mesopores, are accessible from the outside by pore throats with diameters below 2 nm [16]. The existence of all this porosity makes these films have rather low densities and large specific surfaces in contact with the outside, as required for numerous applications, e.g., in sensors, biomedicine or plasmonics, among others [17-26]. This type of morphology has already been investigated for thermochromic applications in the literature, mainly because of their relatively low density and lower effective refractive index [27]. Yet, the existence of a large accessible pore network in the films may represent an even more interesting feature when exposing the film to an oxidative atmosphere, as the uptake of oxygen would not only be limited to the top surface of the film, as it is expected to happen in compact homogeneous films, but rather throughout the whole accessible pore structure. Based on this idea and the results obtained in ref. [11], in this work, the low temperature oxidation of different nanocolumnar amorphous $VO_x$ thin films has been systematically analyzed, aiming at maximizing the formation of monoclinic $VO_2$ in the material. In addition to the structural and chemical characterization of the films, the metal-insulator transition has also been studied by analyzing the optical and electrical modulation properties during the MIT. The obtained results do not only demonstrate the feasibility of this methodology to fabricate thermochromic coatings with relatively good optical properties at temperatures below 300ºC, but also provide interesting clues to understand the importance of porosity and its interplay with the oxidation process for the stabilization of $VO_2$ crystalline domains.

## 2.-Experimental Setup

### *2.1 Materials*

$VO_x$ thin films were deposited at room temperature on doped silicon (1 0 0) wafers (Topsil) diced in 1 x 1 $cm^2$ pieces, on silica (Vidrasa) diced in 1 x 1 $cm^2$, and on a (0001) 99.9% pure $Al_2O_3$ sapphire substrate with 31 mm diameter and a surface roughness lower than 0.5 nm. For MS-OAD, a V target with 3' diameter and 99.95% purity (Testbourne) was used. Ar and $O_2$ were used as plasma gases, with 99.99995% purity (Air Liquide). Prior to the deposition, the substrates were cleaned by conventional procedures in an ultrasonic bath (acetone, isopropanol, and deionized water). Other reagents and solvents were purchased as reagent-grade and used without further purification.

### *2.2.-Deposition and oxidation of nanocolumnar $VO_x$ thin films*

Nanocolumnar and amorphous thin films containing different amounts of O and V were deposited at room temperature by means of MS-OAD in a cylindrical deposition reactor, 75 cm long and radius of 18 cm, with base pressure of 2 x$10^{-4}$ Pa  (see scheme in Figure 1a). A 3 inch diameter V target was used as a cathode, with a substrate holder located 15 cm apart, and aligned with the center of the target. An argon gas flow of 9.3 sccm was pumped into the reactor chamber along with a small flow of oxygen gas of 0.2 sccm that yielded a total pressure of 0.2 Pa. An electromagnetic power generator (Advanced Energy DC Pinnacle Plus) was employed to ignite a DC plasma, setting the value of the power between 150 W and 275 W. The substrate holder was tilted 85º with respect to the target's surface, in a typical MS-OAD configuration, while the deposition time was set to obtain film thicknesses of around 300 nm. Table I contains the list of conditions employed to deposit the coatings studied in this work. A second set of depositions were carried out to grow compact thin films with different [O]/[V] ratios. In this case, the classical geometrical configuration was employed, placing the substrate holder parallel to the target's surface, i.e., a substrate tilt of 0º, and a power of 150 W (see deposition conditions in Table I). In all cases, a pair of substrates consisting of a silicon wafer and a silica plate were utilized. For electrical and Hall effect measurements, an additional set of sapphire substrates (dimensions: 3× 3 cm) were used under the conditions described below.

After the deposition of the $VO_x$ samples, these underwent an oxidation procedure in a furnace at atmospheric pressure using a temperature controller EUROTHERM 2408 (see Figure 1b). Oxygen gas was fed in the furnace chamber, imposing a heating ramp of 5ºC $min^{-1}$, up to a given targeted temperature that was kept constant during 90 min, after which the whole furnace was left to cool down until it reached room temperature (here it is worthy to mention that similar results as those presented in this paper have been obtained when using atmospheric air instead of pure oxygen for the annealing process).

### *2.3.-Characterization of nanocolumnar $VO_x$ thin films*

The morphology of the films was characterized by means of scanning electron microscopy (SEM) with a high-resolution field emission gun microscope model HITACHI-S4800. The microstructure of the layers was also studied by Scanning Transmission Electron Microscopy (STEM): cross-sectional slices of the films deposited on silicon substrates were prepared using the conventional procedure of mechanical polishing followed by argon ion milling to electron transparency. High-angle annular dark field -STEM (HAADF-STEM) micrographs were acquired in a Tecnai G2 F30 S-Twin STEM from FEI, equipped with a HAADF detector from Fischione with a 0.16 nm point resolution.

The [O]/[V] atomic concentration ratio in the films was determined by means of the Rutherford Backscattering Spectroscopy (RBS) and Nuclear Reaction Analysis (NRA) techniques. The setup employed to carry out these experiments was a 3 MV tandem Accelerator at the Centro Nacional de Aceleradores (CNA, Seville, Spain) employing a beam of 2.0 MeV alpha particles for RBS and 0.9 MeV deuteron beam for the NRA, with a 1 mm beam spot diameter, and passivated implanted planar silicon detectors, located at 165$^o$ scattering angle for RBS and 150$^o$ for NRA. In the latter case, a 13 µm thickness filter of aluminized Mylar was placed in front of the detector to avoid scattered deuterons reaching the detector. The spectra have been analyzed using SIMNRA 6.0 code [28]. To ensure a consistent analysis, all reported measurements (morphological, chemical, electrical and optical) have been performed at the central location of the substrate.

Raman spectroscopy analyses of the samples have been carried out using a LabRAM Horiba Jobin Yvon with a green laser beam of 532 nm wavelength. An X-ray investigation was carried out by means of a D8 DISCOVER diffractometer. The device featured a 2D detector VANTEC-500 with a 2 mm capillary, which allows to obtain 2D frames corresponding to each point ranging from 10$^o$ to 60$^o$ 2θ. A copper Kα radiation source with 0.15405 nm wavelength has been used with a step width of 20$^o$, for 1 h per step and tube conditions of 40 kV voltage and 40 mA current. Grazing Incidence X-ray Diffraction measurements were acquired by maintaining a fixed incidence angle of 0.5°.

The optical transmission properties of the films have been analyzed in the wavelength range of 200–2500 nm by means of an ultraviolet-visible spectrophotometer PerkinElmer Lambda 750 S with a 60 mm diameter integrating sphere. In order to characterize the metal-insulator transition, a homemade device consisting of two ceramic heaters with a pierced hole at the center was used to hold the samples during the ultraviolet-visible spectra acquisitions. The sample was heated up to 100$^o$C during optical characterization, with the device connected to a programmable power supply ISOTECH IPS-405 operated in DC mode, and using a thermocouple to read and control the temperature. Finally, the transmittance-versus-temperature thermochromic hysteresis loops were obtained employing the same setup, for a wavelength of 1500 nm in variable steps of temperature ranging from room temperature to 100$^o$C. DC electrical resistivity *vs.* temperature measurements of the furnace oxidized films were performed in a dark environment with a homemade setup. It involved a four-probe system with a van der Pauw geometry in the temperature range of 25–100$^o$C with a ramp of 1$^o$C min$^{-1}$ and then back to 25$^o$C with an equivalent negative ramp. Carrier mobility and carrier concentration were obtained by Hall effect measurements using the same van der Pauw geometry. A constant magnetic field of 0.800 T was applied perpendicular to the sample surface.

In terms of notation and labelling of the samples, in this paper we have chosen the following criterion: as-deposited films have been labelled with the prefix "nano-" or "compact-" depending on whether the structure is nanocolumnar or compact, respectively, followed by "$VO_x$", with x being the measured value of the [O]/[V] ratio in the as-deposited film. For instance, nano-$VO_{1.5}$ refers to an as-deposited nanocolumnar film with a measured value of [O]/[V]= 1.5. Moreover, the films that have undergone an oxidation process also include the value of the oxidation temperature in their label. For instance, the film nano-$VO_{1.5}$(280ºC) refers to a nanocolumnar film with an as-deposited stoichiometry of [O]/[V]=1.5 that has been subjected to oxidation at a temperature of 280ºC (note that the value of the [O]/[V] ratio after the oxidation process is not included in the label).

## 3.-Results and Discussion

### *3.1.-Analysis of the nanocolumnar VOx thin films before and after the oxidation process*

As mentioned in the previous section, a set of nanocolumnar thin films was grown with different [O]/[V] ratios by means of the MS-OAD technique, by varying the value of the DC sputtering power from 150 W to 275 W and maintaining a constant flow of oxygen in the reactor (see Table I). The measured value of the [O]/[V] ratio of these films is depicted in Figure 2, where a clear decreasing trend in the oxygen content with power is obtained, from [O]/[V](150 W)=1.9 to [O]/[V](275 W)=1.3. This systematic change can be explained by the increase of the sputtering rate of V atoms for increasing powers and, thus, of the deposition rate of V atoms. From a morphological point of view, the as-deposited films are alike, as illustrated in Figure 3a-b, where the cross-sectional FESEM images of nano-$VO_{1.9}$ (Figure 3a, top left) and nano-$VO_{1.3}$ (Figure 3b, top left) are shown. Interestingly, and despite the different chemical composition of both films, a similar columnar morphology is apparent in both cases: Tilted ~30º with respect to the substrate normal, with column diameters of ~100 nm and separated by noticeable large intercolumnar mesopores. The origin of such similar morphology stems from the common geometrical oblique angle arrangement in both cases, which is known for mediating surface shadowing mechanisms and the subsequent formation of the tilted nanocolumnar structures. This is apparent in the top view images of these films, also depicted in Figures 3a-b (top right images), where similar nanocolumnar and pore morphologies can be observed. Moreover, the absence of peaks in the Raman analyses, along with the high opacity of these coatings (results not shown), indicate that they are mostly composed of oxygen-deficient amorphous V-O domains.

The as-deposited films were placed in a furnace and exposed to an oxygen atmosphere at different temperatures, as described in the Experimental Setup section. After this oxidation process, the value of the [O]/[V] ratio in the films clearly increased. For illustration purposes, the measured values of [O]/[V] after oxidation at 280ºC are included in Figure 2. There, it is remarkable that the film nano-$VO_{1.9}$ reaches the oxygen saturation value ([O]/[V]=2.5) after oxidation, in agreement with the results in ref. [11], while the remaining samples present post-oxidation stoichiometries between 1.9 and 2.3. Moreover, from a morphological point of view, the oxidation process also causes important structural changes, which is illustrated in Figure 3a-b, where the top and cross-sectional FESEM views of the coatings nano-$VO_{1.9}$ (280ºC) (figure 3a, bottom images) and nano-$VO_{1.3}$ (280ºC) (figure 3b, bottom images) are displayed. There, it is noticeable that the oxidation has caused the widening of the nanocolumns up to the point of even merging at some locations, shrinking the pores and conforming a more compact structure. Furthermore, Figure 3 evidences an overall increase of thickness of about 20% in both cases, despite the fact that these images were taken from the same samples before and after the oxidation process and, approximately, at the very same location (positioning error below 1 mm), finding a shift in thickness from 280 nm to 335 nm in the case of nano-$VO_{1.9}$ and from 310 nm to 375 nm in the case nano-$VO_{1.3}$. We attribute this swelling of the structure to the incorporation of O into the film network, which is known for increasing the associated volume per V atom in the film (a simple calculation based on the standard density of different vanadium oxide materials results in increasing values of volume per V atom in the material, from $\sim 1.4 \times 10^{-2}\ nm^3/V\ atom$ in pure V to $\sim 4.5 \times 10^{-2}\ nm^3/V\ atom$ in $V_2O_5$). This swelling phenomenon would additionally be favored by the relatively high mobility of the species in the $V_2O_5$ domains at such low temperatures (as it will be demonstrated below, this phase appears in all these coatings after oxidation), with a Tamman temperature as low as ~200ºC (for comparison purposes, the Tamman temperature in $VO_2$ amounts to ~847ºC). This swelling phenomenon is further confirmed in Figure 4, where HAADF-STEM images of the tips of the nanocolumns in nano-$VO_{1.9}$ and nano-$VO_{1.9}$(280ºC) are displayed, and where it is clearly shown how the intercolumnar mesopores shrink and a more compact structure is formed after oxidation.

In Figure 5a, the Raman spectra of the nanocolumnar films after oxidation at 280ºC are presented. There, it is apparent that nano-$VO_{1.9}$(280ºC) contains $V_2O_5$ crystalline domains, which is coherent with the measured value [O]/[V]~2.5 reported in Figure 2. Moreover, the spectrum of nano-$VO_{1.7}$ (280ºC) indicates not only the existence of $V_2O_5$, but also of $V_3O_7$ and $VO_2$ crystalline domains in the film. This becomes even more evident for samples nano-$VO_{1.5}$ (280ºC), nano-$VO_{1.4}$ (280ºC) and nano-$VO_{1.3}$ (280ºC) where the peaks associated with $VO_2$ are clearly discernable. Consequently, it has been found that whenever the as-deposited [O]/[V] ratio stays below 1.9, the oxidation process at 280ºC triggers the formation of $VO_2$ along with the $V_2O_5$ and $V_3O_7$ phases. This is also corroborated by XRD analyses of the same films, presented in Figure 5b, where the peaks corresponding to monoclinic $VO_2$, $V_2O_5$ and $V_3O_7$ phases are evident.

In addition to the [O]/[V] ratio in the original (as-deposited) films, the oxidation temperature also plays a key role in the formation of $VO_2$ crystalline domains. In Figure 6a-b, the Raman spectra and the XRD patterns of the film nano-$VO_{1.5}$ after being subjected to an oxidation procedure at temperatures of 260ºC, 280ºC and 300ºC are shown, respectively. There, the peaks associated with $V_2O_5$ are evident in all the cases, while the $V_3O_7$ and $VO_2$ phases are clearly discernable in nano-$VO_{1.5}$ (280ºC), as well as in nano-$VO_{1.5}$ (260ºC) and nano-$VO_{1.5}$ (300ºC), even though they are not as intense in these two latter cases. This implies that nano-$VO_{1.5}$ (280ºC) seems to represent the optimum case in terms of $VO_2$ peak intensity. In addition, and according to Figures 5 and 6, it is remarkable that $VO_2$ in these coatings always coexists with the $V_2O_5$ and the $V_3O_7$ phases, which provides some clues on how the oxygen incorporates into the film network and induces the formation of different crystalline domains, being $V_2O_5$ the only crystalline phase that has been found isolated from the other two. It is noteworthy in this regard that the crystal size determined by the Scherrer equation [29] from the width of the peaks in Figure 5b renders approximate values of ~10 nm for $VO_2$, ~15 nm for $V_3O_7$ and ~25 nm for $V_2O_5$. Later on, in section 3.3, a model on the formation of $VO_2$ at low temperatures that takes into account the film stoichiometry, porosity and oxidation temperature on the formation of $VO_2$ is put forward.

*3.2.- Optical and thermochromic properties*

Figure 7 presents a photograph of the samples deposited on $SiO_2$ substrates after the oxidation process at different temperatures. These films present a brownish color, which is typical of V-O thin films, and show certain transparency, especially the films nano-$VO_{1.9}$(260ºC, 280ºC and 300ºC) and nano-$VO_{1.7}$(260ºC, 280ºC and 300ºC), no matter the oxidation temperature. This agrees with the existence of a large amount of $V_2O_5$ crystalline domains in these cases, as obtained in the XRD/Raman analyses in Figures 5 and 6. Interestingly, the transparency of the remaining samples is highly dependent on the oxidation temperature. They are slightly opaque when the temperature is 260ºC, and increasingly transparent when the temperature is 280ºC and 300ºC, suggesting that a certain amount of (oxygen deficient) amorphous $VO_x$ regions survive the oxidation process, which are progressively removed as the oxidation temperature increases.

In Figure 8, the UV-vis spectra of the films depicted in Figure 7 are presented when measured both at room temperature and at 100ºC. As expected, the films nano-$VO_{1.9}$ (260ºC), nano-$VO_{1.9}$ (280ºC), nano-$VO_{1.9}$ (300ºC) show typical spectra of $V_2O_5$ as well as no thermochromic behavior. This agrees with the XRD/Raman analyses in Figures 5 and 6, and with the lack of any $VO_2$ signal in these cases. Remarkably, sample nano-$VO_{1.5}$ (280ºC) possesses a transmittance spectrum quite similar to that reported in literature for (non-doped) $VO_2$ thin films [30, 31, 32], characterized by a clear thermochromic transition in the IR region from ~60% at room temperature to ~10% at 100ºC, with a transmittance in the visible part of the spectrum of ~40%, no matter the environmental temperature. This agrees with the Raman and XRD results in Figures 5 and 6 and the relevant amount of $VO_2$ crystal domains detected for these samples. For the rest of the cases, the optical modulation is coherent with the existence of $VO_2$ crystalline domains, as measured in Figures 5 and 6, along with amorphous oxygen-deficient $VO_x$ that hinders the overall optical transparency of the coatings. In this regard, and according to Figure 7, the amount of amorphous $VO_x$ after oxidation seems to diminish for increasing values of the as-deposited [O]/[V] ratio and the oxidation temperature. This result indicates the delicate interplay between as-deposited nanocolumnar film stoichiometry and oxidation temperature, not only for the formation of the different crystalline phases, namely $V_2O_5$, $V_3O_7$ and $VO_2$, but also for the removal of the oxygen deficient amorphous $VO_x$ regions, this latter with a strong influence on the overall transparency of the film.

Given the results presented in Figure 8, we have chosen the film nano-$VO_{1.5}$ (280ºC) as the best case in terms of thermochromic efficiency, and to further analyze its optical and electrical modulation capabilities during the MIT. Figure 9 showcases the changes in transmittance at a wavelength of 1500 nm when varying the temperature of nano-$VO_{1.5}$ (280ºC) from room temperature to 100ºC and back. A typical hysteresis loop corresponding to a metal-insulator transition is reproduced, where a transmittance drop of ~50% (from ~65% to ~15%) is found, with the centroid of the hysteresis loop at a transition temperature of $T_f$ = 50.3ºC. To evaluate the thermochromic efficiency of this coating, different standard quantities have been calculated, such as $T_{lum}$ that quantifies the visible light transmitted by the coating, and is used as an indicator of luminosity (see appendix I in ref. [11] for its definition). The value of this parameter is 26.5% (at room temperature) and 23.1% (at 100ºC). Moreover, the solar modulation, $\Delta T_{sol}$, taken as a reference of the radiation not transmitted by the film due to the thermochromic transition, has a value of 12.5%. This means that this coating possesses a high enough optical modulation as to be employed for practical applications (the general consensus for this matter is $\Delta T_{sol} > 10\%$ for a proper adaptation to different climate changes [33, 34]). In this regard, the only constrain for its direct use is its poor optical transparency in the visible range, which can be improved by means of well stablished strategies already available in the literature, e.g., by using doping techniques [31]. The electrical modulation features of Nano-$VO_{1.5}$(280ºC) during the MIT were also assessed by means of the van der Pauw method and Hall effect as a function of temperature (note that the substrate employed to support the coating for this study was sapphire, with larger size than the silica substrate). This film underwent a significant temperature-induced transition during heating, where the DC electrical resistivity abruptly changed more than two orders of magnitude from $10^{-1}$ to $10^{-3}$ $\Omega$·m when varying the temperature from 25 to 100ºC (Figure 10a), being this drop characteristic of $VO_2$ compounds [35]. Interestingly, the measured hysteresis loop depicts a similar sharp increase of resistivity when the temperature surpassed ~53ºC, i.e., 23ºC below the nominal value of the transition at 68ºC for pure $VO_2$ phases [36]. Since XRD and Raman analyses in Figures 5-6 both showed the occurrence of $V_3O_7$, $V_2O_5$ phases in nano-$VO_{1.5}$(280ºC), a smaller variation of electrical resistivity compared to that of pure or single crystalline $VO_2$ material would be expected. However, jumps of resistivity, as well as the width and position of the measured hysteresis (see Figure 10a) support that preparation conditions of nano-$VO_{1.5}$(280ºC) favor the formation of a significant amount of $VO_2$ domains in detriment of other phases and/or oxygen-deficient regions. Hall effect measurements on nano-$VO_{1.5}$(280ºC) show that electrons are the majority carriers whatever the temperature range, revealing the n-type nature of this film. Electron concentration and mobility *vs.* temperature also show strong variations and hysteresis loops, as shown in Figures 10b and 10c, respectively. Upon increasing temperature, concentration increases from $2.3\times10^{19}$ $m^{-3}$ at 25ºC up to $2.0\times10^{24}$ $m^{-3}$ at 100ºC. For these same temperatures, mobility exhibits a reverse evolution reducing from $4.9\times10^{-2}$ $m^2V^{-1}s^{-1}$ down to $8.3\times10^{-4}$ $m^2V^{-1}s^{-1}$, which shows that the drop of resistivity can be mainly assigned to the electron concentration change, in agreement with results in the literature [37-40]. During the heating stage and up to 70ºC, it is noteworthy that the concentration increases continuously with temperature (whereas mobility remains in the same order of magnitude, i.e., in-between 2-$5\times10^{-2}$ $m^2V^{-1}s^{-1}$), which defines a semiconducting-like behavior. Compared to optical transmittance *vs.* temperature measurements performed at a wavelength of 1500 nm (Figure 9), it is important to remark that the transition of electron transport properties occurs at a higher temperature, with a centroid of the resistivity hysteresis at about ~62ºC, which is significantly higher than that determined from the optical transmittance (50.3ºC). Such difference has also been reported without discussing this effect [41-44]. One may suggest that the higher temperature values found for the resistivity hysteresis can be assigned to the difference in the interactions of the $VO_2$ phase with electrons and photons. Optical transmittance is sensitive to the change of local dielectric properties, whereas resistivity depends on electrical current travelling through the different phases and grains, and so on the distance between the electrodes. As a result, current represents a macroscopic probing of the transition and is less sensitive to mean local dielectric variations. Yet, this discrepancy can also be attributed to the different size and nature of the substrate employed for the electrical characterization or to small inhomogeneities in the coating that might affect the measurements.

### *3.3.- Low temperature formation of $VO_2$ in porous nanocolumnar thin films*

The results above, by which small $VO_2$ domains have been crystallized in its monoclinic phase at temperatures below 300ºC, contrast with the nominal value of the crystallization temperature of $VO_2$ in the bulk form well above 500ºC. This means that, in our case, the oxidation process must act as an excitation source that favors the low temperature crystallization of $VO_2$, as it was also concluded in refs. [10, 11]. Moreover, it has been demonstrated that the value of the original (as-deposited) film stoichiometry and the oxidation temperature play key roles in the formation of the different crystalline phases as well as in the survival of oxygen-deficient amorphous $VO_x$ in the material after oxidation. Yet, the influence of the film porosity has still not been specifically addressed, as this feature was rather similar in all the as-deposited films presented above. Keeping this purpose in mind, additional experiments have been carried out on a set of compact layers deposited using the classical magnetron sputtering arrangement under the conditions listed in Table I. These films were deposited with a similar thickness as the nanocolumnar ones (~300 nm) and with different [O]/[V] ratios, labelled as compact-V, compact-$VO_{0.4}$, compact-$VO_{0.9}$ and compact-$VO_{1.5}$. The as-deposited [O]/[V] ratio in these films was measured using the same methodology (RBS/NRA), finding the values 0.0, 0.4, 0.9 and 1.5, respectively. After deposition, all of them were opaque, showed a compact nanostructure, and, attending to the absence of peaks in the Raman and the XRD analyses, did not show any sign of crystalline phases (results not shown), indicating that, similarly to the as-deposited nanocolumnar films, they are formed by oxygen-deficient amorphous $VO_x$.

The compact layers were subjected to the same oxidation process as the nanocolumnar ones at a temperature of 280ºC, finding out that all of them still remained compact (as an example, the cross-sectional SEM image of compact-$VO_{1.5}$(280ºC) is depicted in Figure 11a) and opaque, pointing out the low efficiency of the low temperature oxidation of this films and the survival of a large amount of amorphous $VO_x$. Yet, the Raman analysis of these coatings (see Figure 11b) reveals the formation of $V_2O_5$ crystal domains in all cases, in agreement with the results obtained with the nanocolumnar films (see figure 5 for instance). Remarkably, the crystalline phases $VO_2$ and $V_3O_7$ did not appear in compact-$VO_{1.5}$ (280ºC), which represent a clear difference with nano-$VO_{1.5}$(280ºC). However, there are traces of $VO_2$ in the samples compact-V(280ºC), compact-$VO_{0.4}$(280ºC) and compact-$VO_{0.9}$(280ºC), demonstrating that the proposed methodology does trigger the crystallization of the $V_2O_5$ and $VO_2$ phases in compact thin films at low temperatures. Yet, it is remarkable the absence of $V_3O_7$ in the compact cases, as this phase was always present whenever $VO_2$ was formed in the nanocolumnar films. Moreover, there is no clear indication of any variation in film thickness associated with the oxidation process at 280ºC: this feature, along with the lack of transparency of these films, suggests that only a small portion of material on the film surface is affected, being the oxidation and the associated swelling phenomenon restricted to this very shallow region. Remarkably, these two features (structural swelling and transparency) are evident whenever a compact film is efficiently oxidized. For instance, in Figure 11c the cross-sectional SEM image of a pure V thin film with original thickness of 350 nm after being subjected to an oxidation process at a much higher temperature (600ºC) is presented for illustration purposes. In this case, the film was not only transparent after oxidation (with composition $V_2O_5$), but also showed clear indications of swelling, with an increase of film thickness from ~350 nm to ~830 nm that even caused the direct delamination of the coating at some locations. Consequently, and based on the results above, a basic scheme of the oxidation process of compact films at low temperatures is presented in Figure 12a: there, it is proposed that the low temperature oxidation only affects a specific shallow region in contact with the oxygen atmosphere, where the (highly mobile) $V_2O_5$ phase and the $VO_2$ crystallites are formed, limiting the diffusion of oxygen towards the inside of the material, which remains unaffected.

Based on the idea presented in Figure 12a, we have also tentatively put forward a basic scheme that rationalizes the main results obtained in this paper on nanocolumnar films, which is shown in Figure 12b. There, the as-deposited film is depicted as an array of large and tilted amorphous $VO_x$ nanocolumns separated by large intercolumnar mesopores that penetrate into the nanocolumns by means of large and elongated micropores (please note that the size of the pores is magnified for clarity purposes). In this way, when the film is exposed to the oxygen atmosphere at low temperatures, the whole surface of the nanocolumns (including that of the accessible pores) becomes oxidized, resulting in the formation of $V_2O_5$ and $VO_2$, just like in the compact case (figure 12a). Following an analogous reasoning, this oxide layer likely limits the diffusion of oxygen towards the inside of the nanocolumns, not affecting the $VO_x$ material inside, in a process that is also enhanced by the swelling of the material and the shrinkage/disappearance of the meso- and nanopores. Consequently, and according to this scheme, the resulting film would contain not only oxygen-deficient amorphous $VO_x$ inside the nanocolumns, but also crystallites of $V_2O_5$ and $VO_2$. In this way, when the original film stoichiometry is high enough (in our case when the as-deposited [O]/[V] ratio is above 1.9) or the oxidation temperature is 300ºC or above, it has been found that the oxidation process becomes efficient enough to oxidize the amorphous $VO_x$ material in the nanocolumns and saturate the structure with oxygen, resulting in the formation of a pure $V_2O_5$ thin film. Additionally, it has also been found that there is an optimum value of the original film stoichiometry/oxidation temperature (in our case when [O]/[V]~1.5 and an oxidation temperature of 280ºC) at which the delicate balance between the adsorption of oxygen on the surface of the pores, the swelling of the material and the shrinkage of the pores, as well as the limitation of the oxygen diffusion towards the inside of the film, produces the removal of most amorphous $VO_x$ while preserving the $VO_2$ crystallites in the structure, which clearly improves the optical transparency of the coatings. Finally, it is also worth mentioning that the schemes in Figures 12a and 12b do not account for the formation of the $V_3O_7$ phase, which is absent in the compact films and that only appears in the nanocolumnar cases whenever $VO_2$ is formed.

Based on the results above, it is demonstrated that the use of nanocolumnar films provides several key advantages in comparison with compact films when subjected to a low temperature oxidation process to form $VO_2$ crystallites, such as i) the nanocolumnar films possess a much higher specific surface that makes the adsorption of oxygen much more efficient (promoting the formation of a larger amount of $V_2O_5$ and $VO_2$ and the removal of a larger amount of undesired opaque amorphous $VO_x$), and ii) the possibility to accommodate the swelling of the material into the structure thanks to the existence of large empty voids that serve to release stress, and that would cause the direct delamination of the coating should it be compact. Finally, it is noteworthy that the features of the film nano-$VO_{1.5}$(280ºC) were stable for more than one year, depicting the same modulation capabilities and structural features than just after its fabrication, despite the fact that it was exposed to the atmosphere during all this time and that it was not protected by any cap layer or by any other means. This implies that both the formation of $V_2O_5/VO_2$ on the accessible surface of the film and the swelling of the material, are enough to limit the diffusion of oxygen and avoid its eventual transformation into the non-thermochromic highest oxidized phase ($V_2O_5$) even after such a long time.

## 4.- Conclusions

From the previous results and discussion, the first conclusion of this work is that there is a window of conditions to prepare $VO_2$-containing thin films at temperatures below 300ºC. The found conditions require the use of porous $VO_x$ thin films as precursor and a precise control of both, the porosity of the films and the [O]/[V] ratio prior to the oxidation. In this regard, the use of the well-established MS-OAD technique has proven to be quite straightforward for a precise control of these features. In particular, the feasibility to produce $VO_2$ at relatively low temperatures (280ºC) using industrially scalable methods as MS-OAD is demonstrated by the low temperature oxidation of an amorphous nanocolumnar $VO_x$ thin film. Actually, the fact that similar results can be obtained using either pure oxygen or an air flow for the low temperature oxidation process suggests that the purity of the oxidizing atmosphere is not critical to achieve the thermochromic property at low temperatures. This evidence, together with the widespread use of the magnetron sputtering technique in the industry support the use of the proposed method as a basis to fabricate thermochromic $VO_2$ thin films at temperatures below 300ºC in mass production facilities.

In this work we have obtained final thin films with variable transparency and thermocromic capabilities. In particular, we have produced a relatively good thermochromic response as proved by the achievement of optimum values of $T_{lum}$ around 25% (a parameter that is usually employed as an indicator of luminosity) for non-doped $VO_2$ thin films and $\Delta T_{sol}$ of 12.5%, this latter taken as a reference of the radiation modulation features. For these samples, resistivity *vs.* temperature also exhibited a hysteresis loop with a variation in the electrical properties of more than 2 orders of magnitude. Hall effect measurements revealed the n-type nature of the films with a reverse evolution of electron concentration and mobility, resulting in abrupt changes of resistivity at the transition temperature that have been mainly assigned to electron concentration changes. Based on a thorough characterization analysis of the films using Raman, XRD, RBS/NRA and electron microscopy tools, a model to explain the main quantities mediating the low temperature formation of $VO_2$ has been proposed. Basically, it takes into account the swelling of the $VO_x$ structure upon the incorporation of oxygen into the network, the low temperature formation of $V_2O_5$ and $VO_2$ crystallites on the surface of the film in contact with the oxygen atmosphere, and the limitation of the oxygen diffusion towards the inside of the material due to the formation of these phases. In this way, the precise control of the initial porosity of the sputtered films ant their actual [O]/[V] ratio have resulted key for a precise control of the final thin film morphology and thermochromic modulation properties.

## CRediT authorship contribution statement

**Hiedra Acosta-Rivera**: Investigation, Visualization. **Victor Rico**: Investigation, Methodology, Supervision. **Francisco J. Ferrer**: Formal analysis, Investigation. **Teresa C. Rojas**: Formal analysis, Investigation, Visualization. **Rafael Alvarez**: Conceptualization, Project administration, Visualization, Writing – review & editing. **Nicolas Martin:**. Investigation, Visualization, Writing – review & editing. **Agustín R. Gonzalez-Elipe**: Conceptualization, Supervision, Writing – review & editing. **Alberto Palmero**: Conceptualization, Funding acquisition, Project administration, Supervision, Writing – original draft, Writing – review & editing.

## Declaration of competing interest

The authors declare that they have no known competing financial interests or personal relationships that could have appeared to influence the work reported in this paper.

Acknowledgements- The authors acknowledge the projects PID2020-112620GB-I00 and PID2021-123879OB-C21 funded by MCIN/AEI/10.13039/501100011033, and the intramural project

202560E034 funded by the Spanish Council of Research. We thank the projects PID2022-143120OB-I00, TED2021-130916B-I00 funded by MCIN/AEI/10.13039/501100011033/FEDER, UE and by "ERDF (FEDER) A way of making Europe". We thank the financial support from the CSIC Interdisciplinary Thematic Platforms (PTI +) "Sustainable Plastics towards a Circular Economy" (PTI-Susplast +) and "Transición Energética Sostenible" (PTI-TRANSENER+) SOUNDofICE H2020-FETOPEN-2018-2019-2020-01 call GA nº: 899352.

Table I

| #Sample | Power (W) | Deposition time (min) | $O_2$ flow (sccm) | Substrate tilt angle (º) | [O]/[V] (± 0.1) |
|---|---|---|---|---|---|
| nano-$VO_{1.9}$ | 150 | 105 | 0.2 | 85 | 1.9 |
| nano-$VO_{1.7}$ | 200 | 93 | 0.2 | 85 | 1.7 |
| nano-$VO_{1.5}$ | 225 | 82 | 0.2 | 85 | 1.5 |
| nano-$VO_{1.4}$ | 250 | 75 | 0.2 | 85 | 1.4 |
| nano-$VO_{1.3}$ | 275 | 68 | 0.2 | 85 | 1.3 |
| compact-V | 150 | 60 | 0 | 0 | 0 |
| compact-$VO_{0.4}$ | 150 | 60 | 0.2 | 0 | 0.4 |
| compact-$VO_{0.5}$ | 150 | 60 | 0.5 | 0 | 0.5 |
| compact-$VO_{0.9}$ | 150 | 60 | 0.7 | 0 | 0.9 |
| compact-$VO_{1.5}$ | 150 | 60 | 1.0 | 0 | 1.5 |

Table Caption. - List of samples analyzed in this paper and deposition conditions.

## Figure Caption

Figure 1.- a) Scheme of the magnetron sputtering reactor employed to grow the nanocolumnar films arranged according to the Oblique Angle configuration. b) Scheme of the furnace employed to oxidize the coatings.

Figure 2.- Measured values of the atomic concentration ratio [O]/[V] in each nanocolumnar film before and after the oxidation process at 280ºC.

Figure 3.- a) Cross-sectional and top view SEM images of the film nano-$VO_{1.9}$ before and after oxidation at 280ºC. b) a) Cross-sectional and top view SEM image of the film nano-$VO_{1.3}$ before and after oxidation at 280ºC.

Figure 4.- STEM-HAADF image of the tips of the nanocolumns in the film nano-$VO_{1.9}$ before (a) and after (b) being subjected to the oxidation process at 280ºC.

Figure 5.- Raman (a) and XRD (b) analyses of the different nanocolumnar films after being subjected to the oxidation process at 280ºC.

Figure 6.- Raman (a) and XRD (b) analyses of the film nano-$VO_{1.5}$ after being subjected to the oxidation process at temperatures of 260ºC, 280ºC and 300ºC.

Figure 7.- Photograph of the nanocolumnar films grown on $SiO_2$ with as-deposited stoichiometry ranging from 1.3 to 1.9 after being subjected to the oxidation process at temperatures of 260ºC, 280ºC and 300ºC.

Figure 8.- Optical transmittance spectra of the nanocolumnar coatings after being subjected to the oxidation process of 260ºC, 280ºC and 300ºC. Results measured at room temperature (black) and at 100ºC (red) are included.

Figure 9.- Hysteresis loop corresponding to the changes in the optical transmittance of the film nano-$VO_{1.5}$(280ºC) measured at a wavelength of 1500 nm as a function of the environmental temperature after a single heating/cooling cycle from 25ºC to 100ºC and back. The quantities $T_{lum}, \Delta T_{sol}$ and $\Delta T_{IR}$ (see main text for their definition) are included along with the transition temperature (centroid of the loop curve), $T_f$.

Figure 10.- Electrical resistivity(a), carrier concentration(b) and mobility(c) as a function of the environmental temperature of the film nano-$VO_{1.5}$(280ºC). A single heating-cooling cycle was applied, which corresponds to a heating from 25ºC to 100ºC then a cooling back to 25ºC.

Figure 11.- a) Cross-sectional image of the film compact-$VO_{1.5}$(280ºC). b) Raman analysis of the compact films after oxidation at a temperature of 280ºC. c) Cross-sectional image of a compact metallic V thin film with as-deposited thickness of 350 nm after being subjected to an oxidation process at 600ºC.

Figure 12.- Scheme of the main mechanism during the low temperature oxidation of the compact (a) and nanocolumnar (b) coatings. Note that the pores in the nanocolumnar case are magnified for clarity purposes.

Figure 1

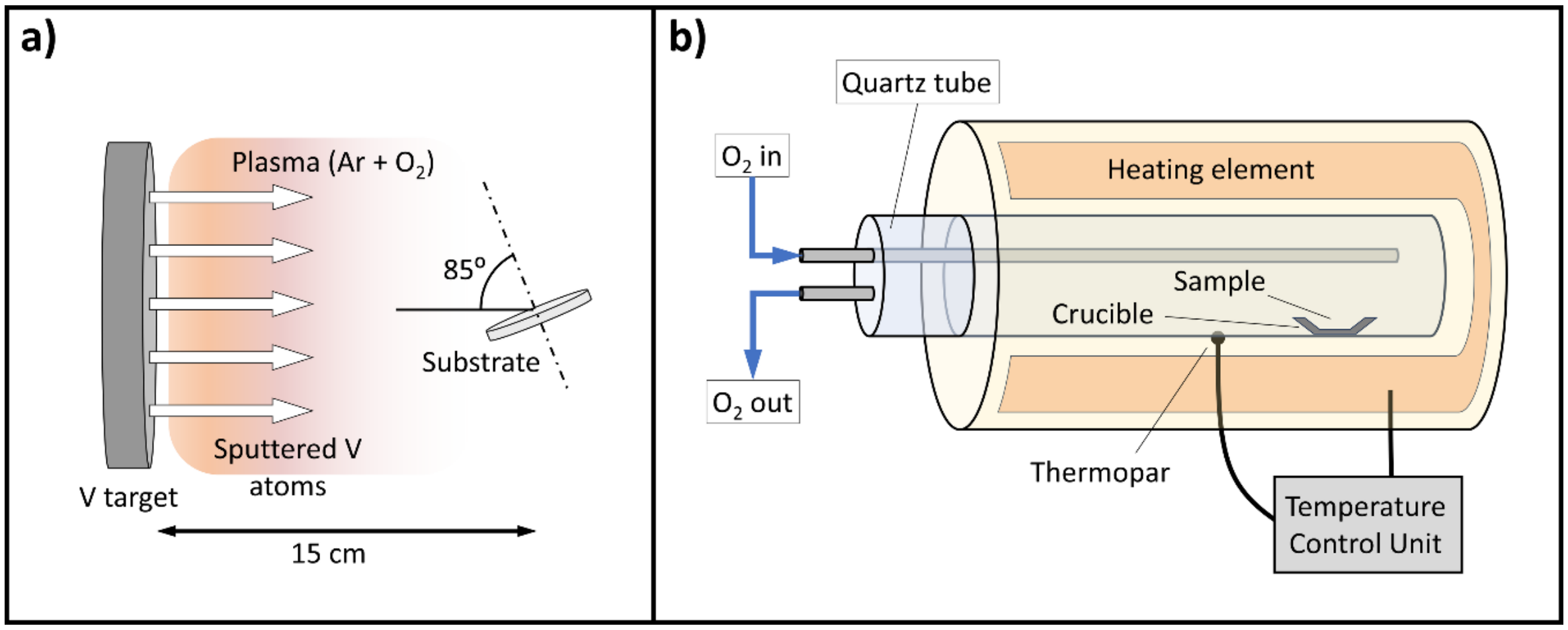

Figure 2

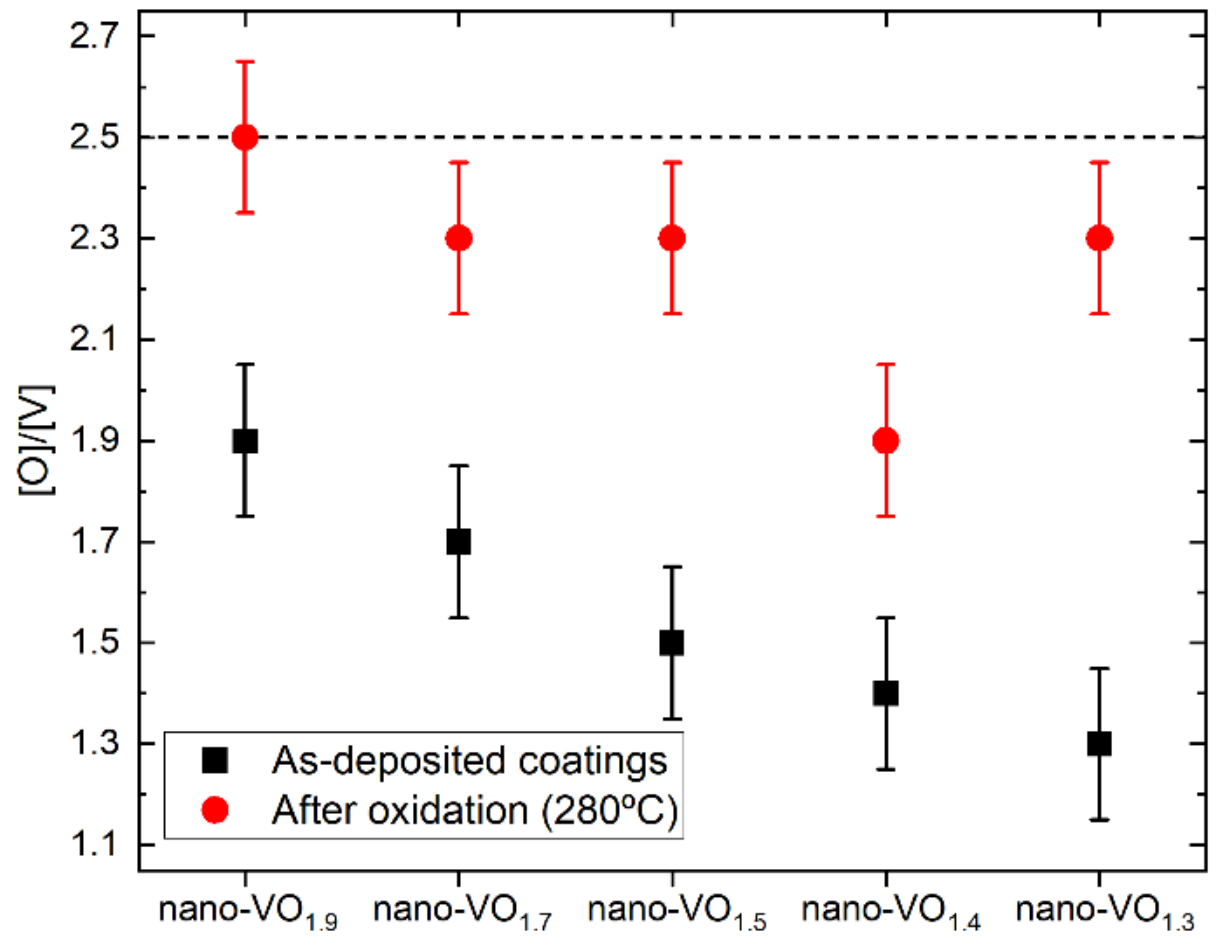

Figure 3

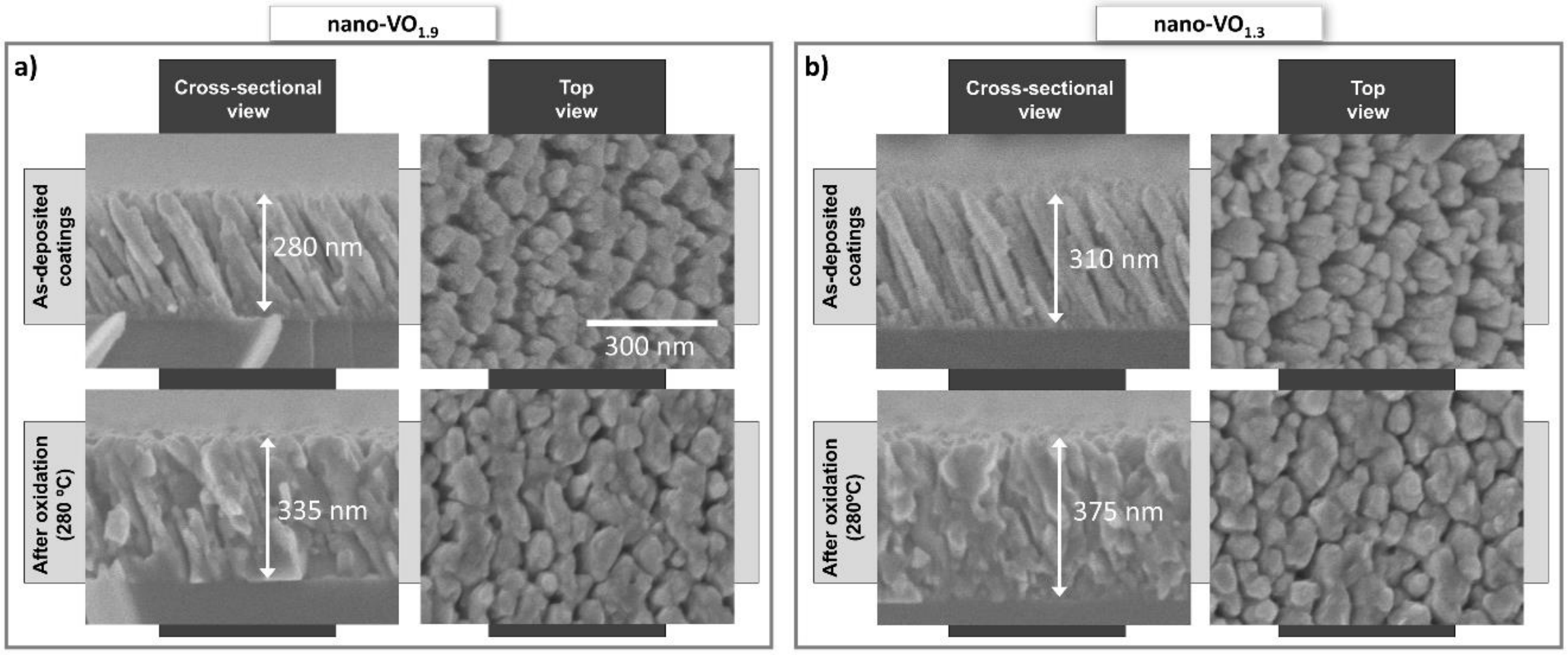

Figure 4

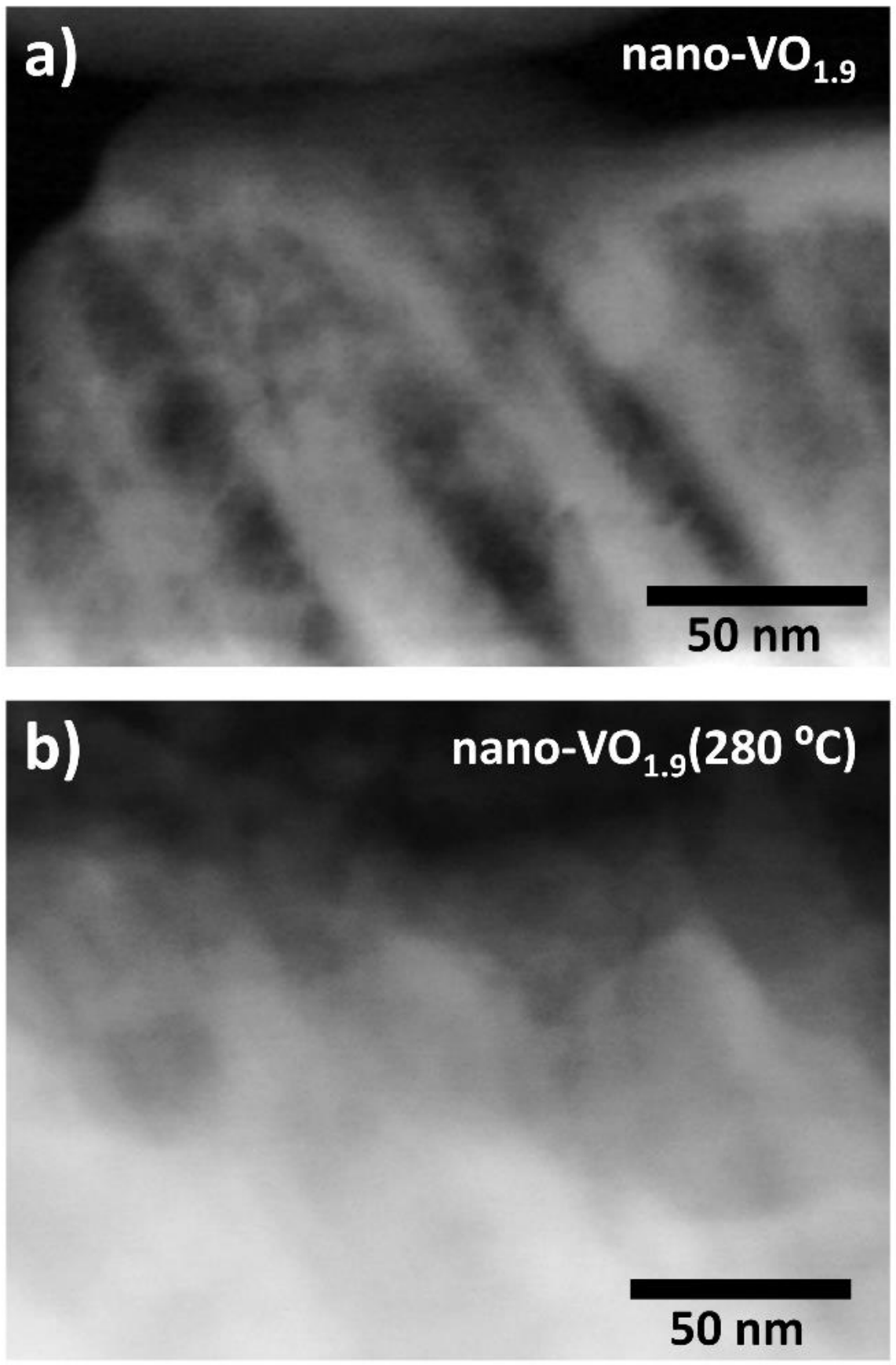

Figure 5

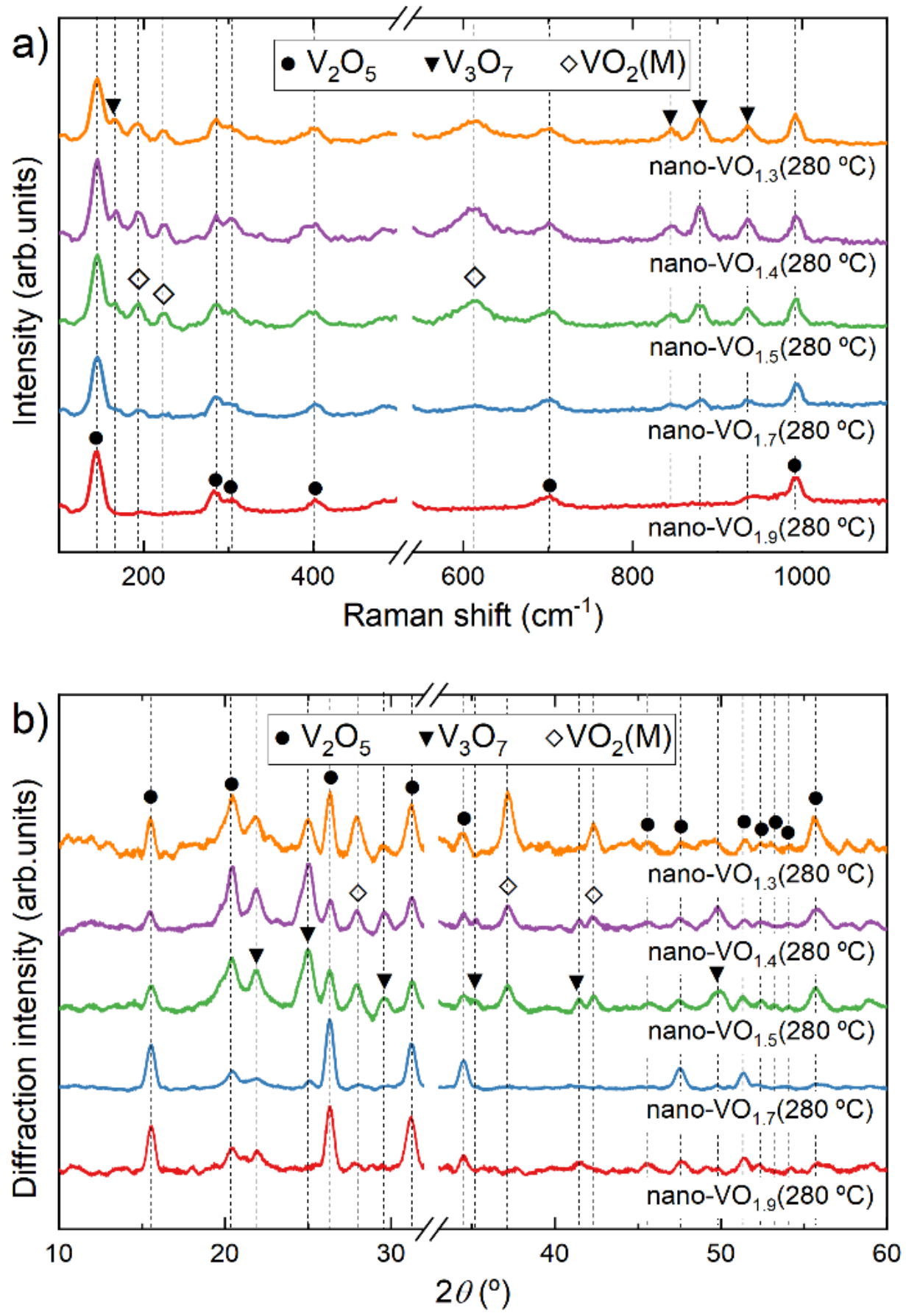

Figure 6

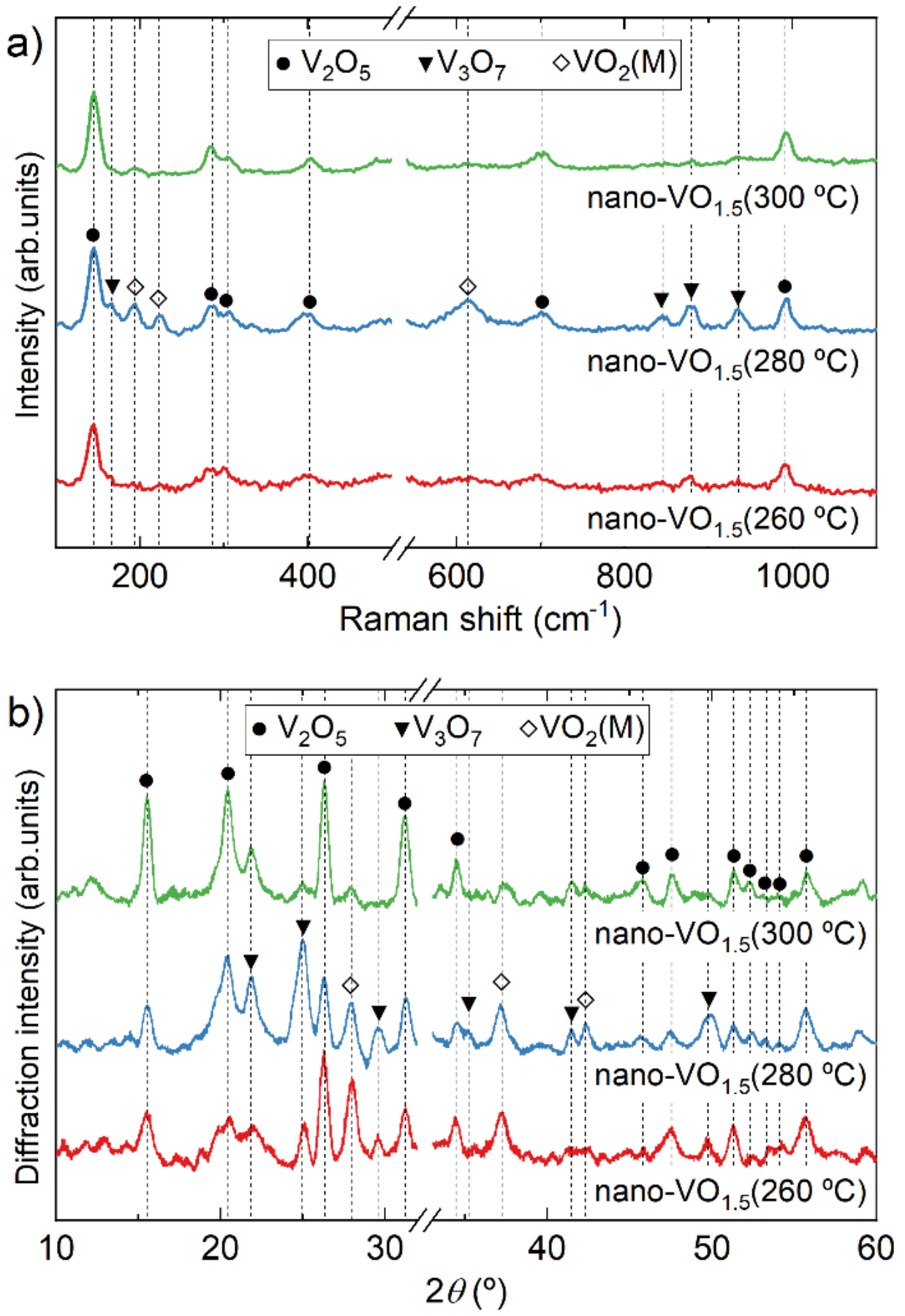

Figure 7

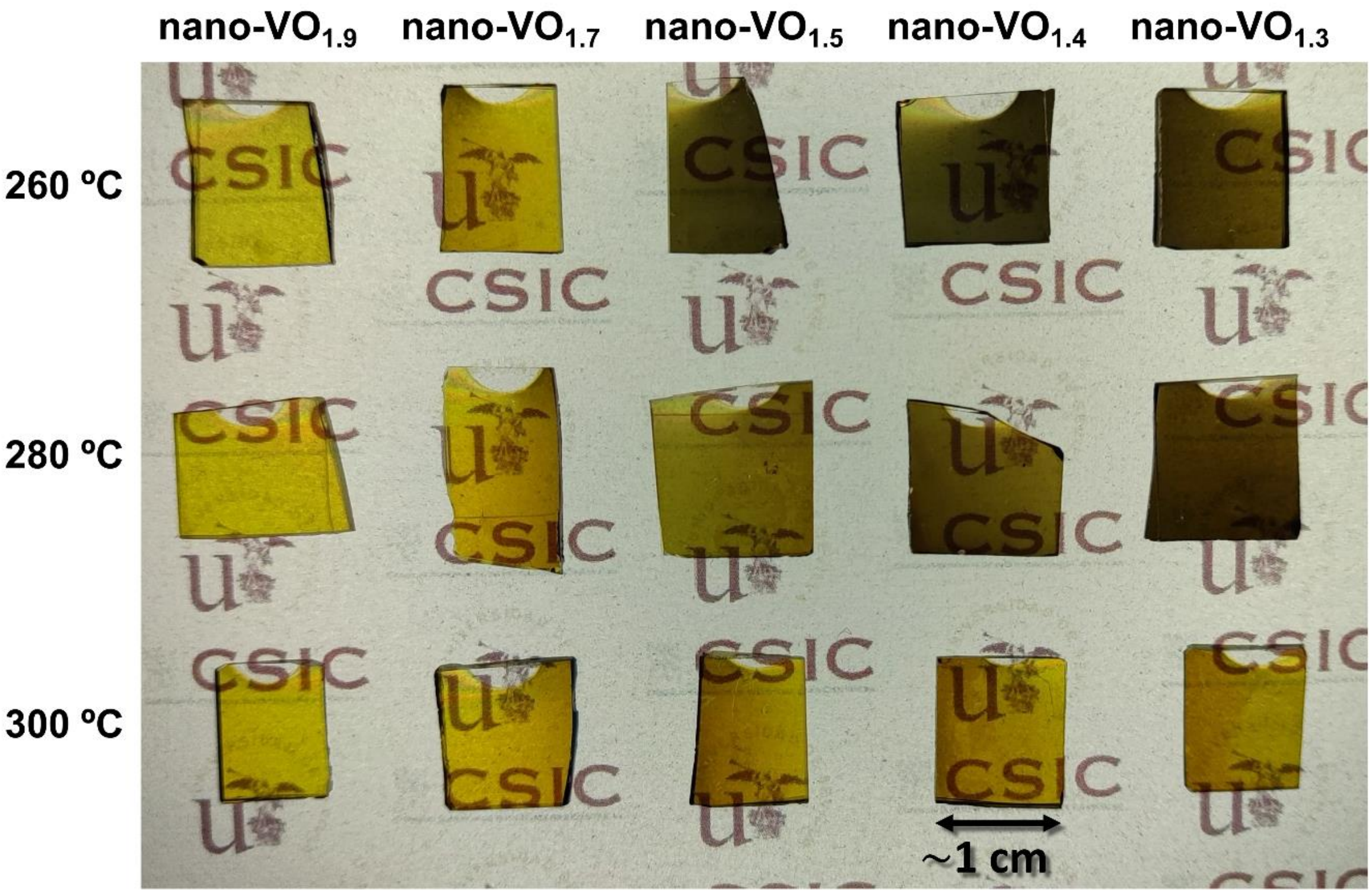

Figure 8

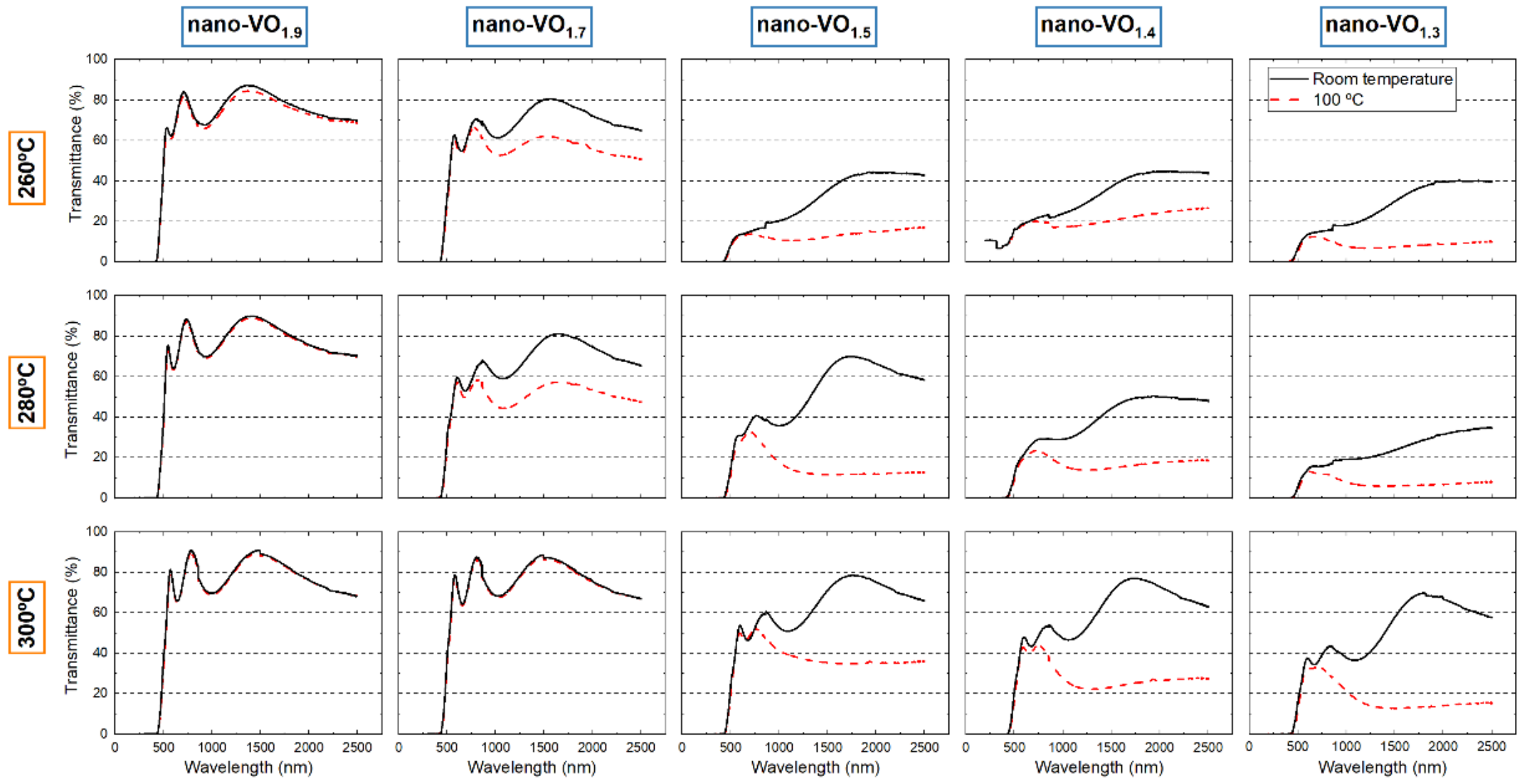

Figure 9

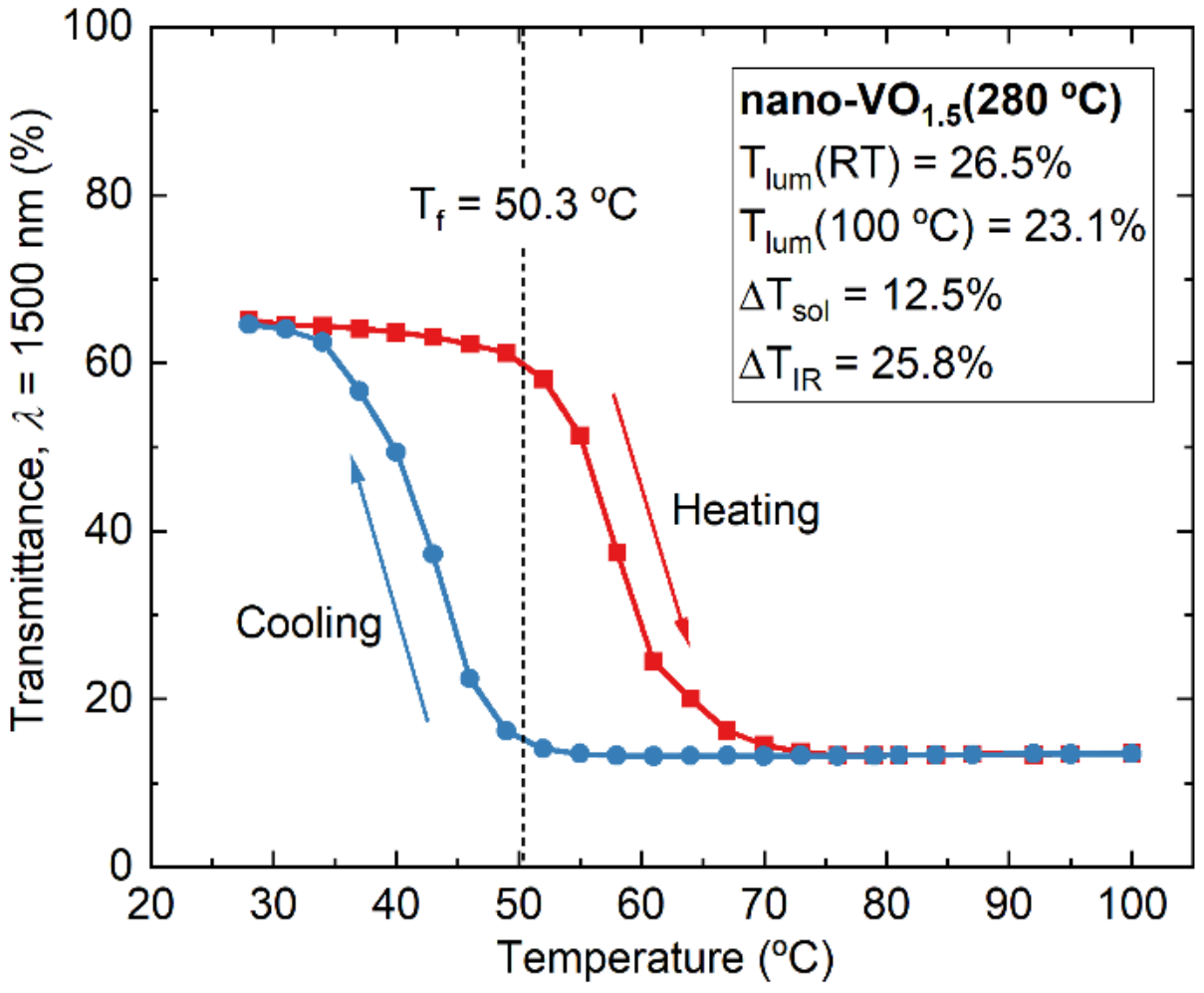

Figure 10

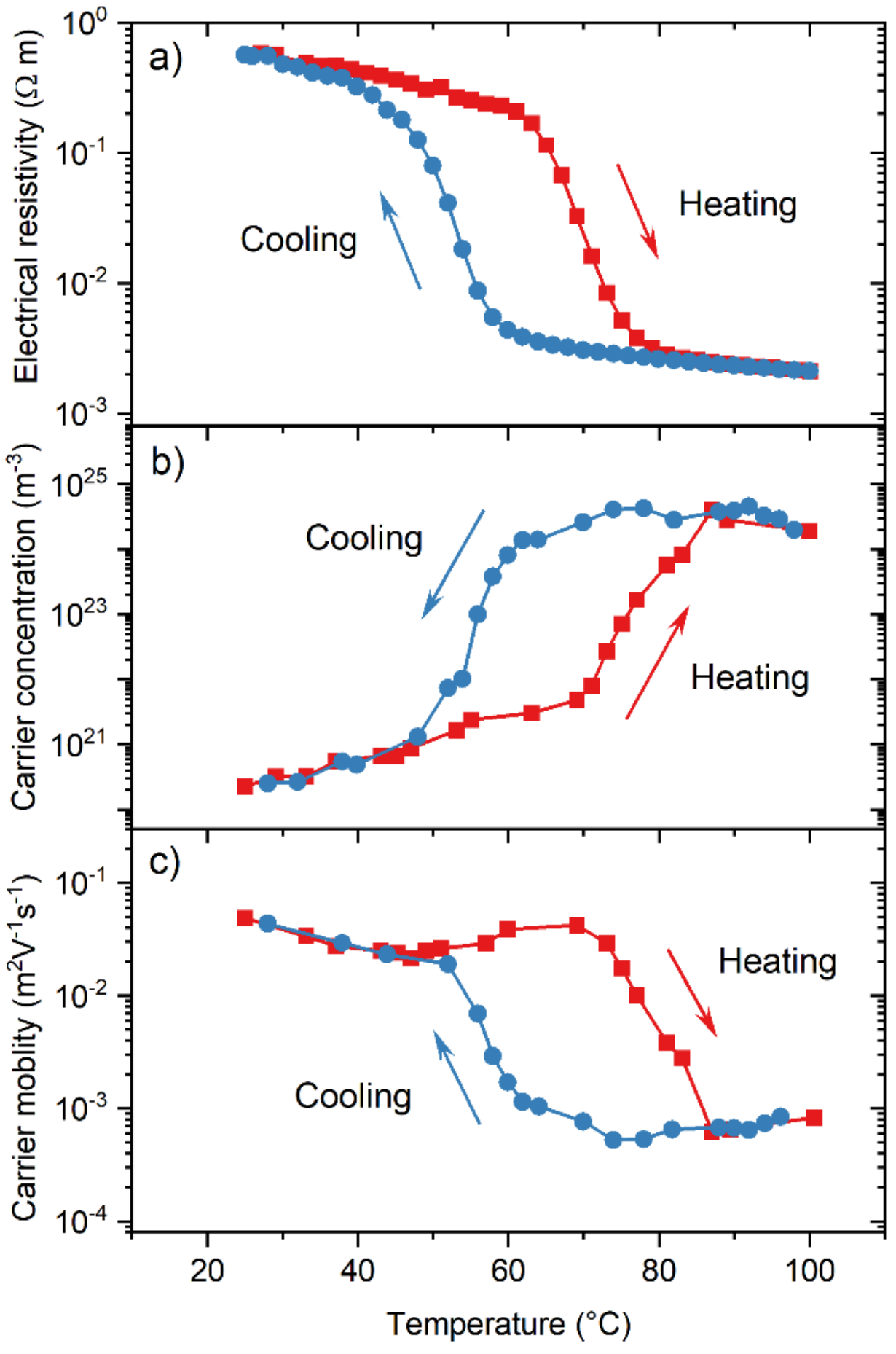

Figure 11

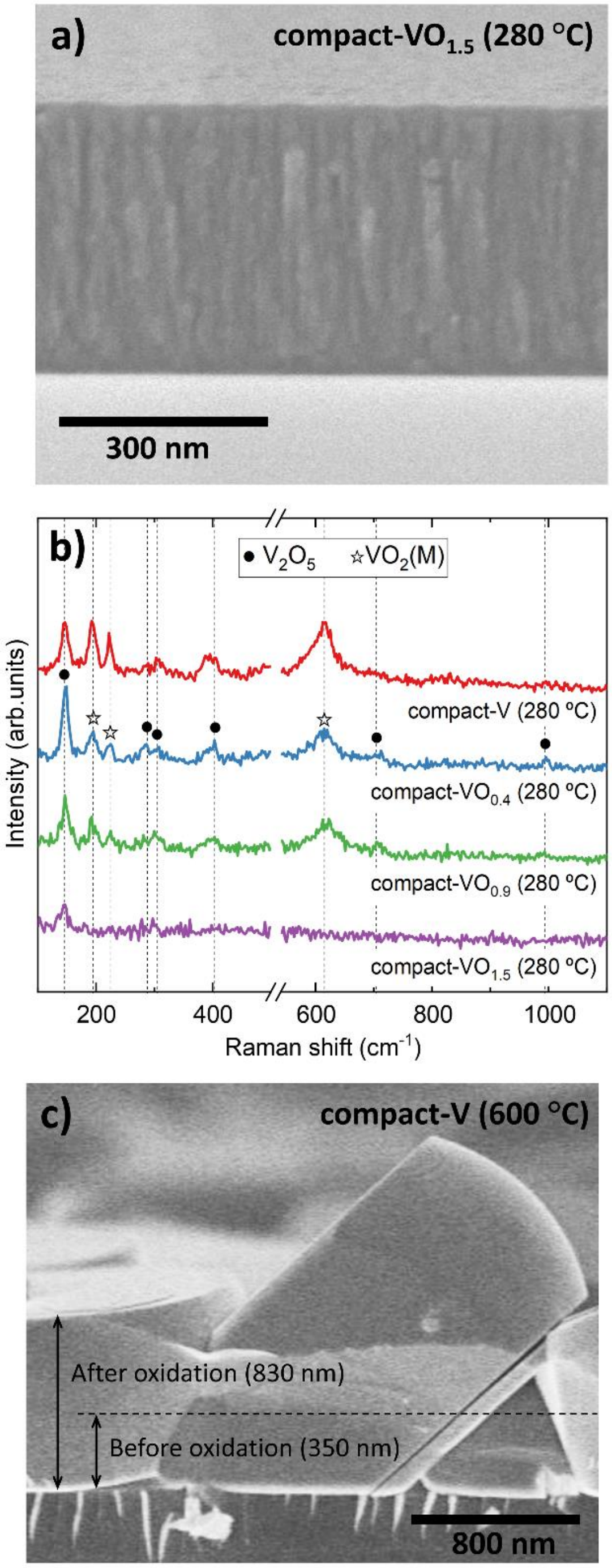

Figure 12

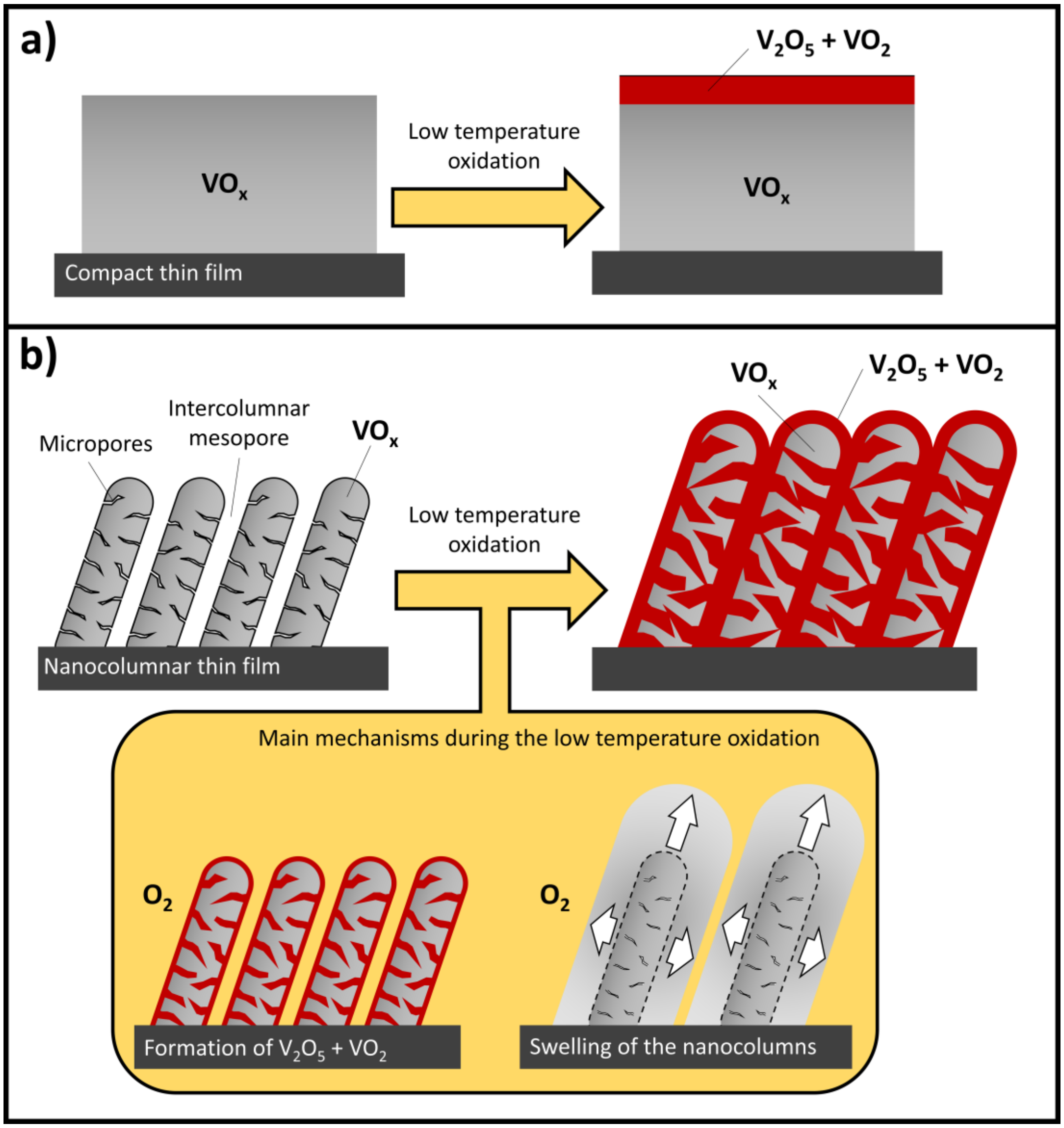